\documentclass[prb,twocolumn,showpacs]{revtex4}
\usepackage{graphicx}
\usepackage{amsmath}
\usepackage{amssymb}
\usepackage{epstopdf}
\usepackage{slashed}

\begin{document}
\title{Models of Strong Interaction in Flat-Band Graphene Nanoribbons: Magnetic Quantum Crystals}
\author{Hao Wang and V. W. Scarola}
\affiliation{Physics Department, Virginia Tech, Blacksburg, Virginia 24061, USA}

\begin{abstract}
Graphene based nanostructures exhibit flat electronic energy bands
in their single-particle spectrum.  We consider interacting
electrons in flat bands of zig-zag nanoribbons.  We present a
protocol for flat-band projection that yields interaction-only
tight-binding models. We argue that, at low densities, flat bands
can delocalize single-particle basis states to support ferromagnetic
quantum crystal ground states.
\end{abstract}


\maketitle

\section{Introduction}

Graphene based structures offer unique opportunities to engineer
electronic band structure by shape alone.
\cite{novoselov:2004,castroneto:2009}  Infinite graphene sheets
exhibit a conic spectrum but finite sized graphene nanostructures
yield a surprisingly broad array of interesting band features.  A
subset of graphene nanostructures reveal flat bands.  Theoretical
work shows that flat bands can be found, e.g., at the edges of
two-dimensional graphene, \cite{nakada:1996} in one-dimensional
graphene nanoribbons, \cite{nakada:1996,lin:2009,potasz:2010}
hydrogenated graphene nanoribbons, \cite{kusakabe:2003} graphene
dots, \cite{ezawa:2007} and graphene antidots. \cite{vanevic:2009}

Electrons in flat kinetic energy bands pose challenging
theoretical problems. The absence of any dispersion leaves the
Coulomb interaction to govern the low energy physics. Many common
approximations fail in the extreme flat-band limit.  A single flat
band cannot lead to intra-band screening as in ordinary Fermi
liquids, e.g., two-dimensional graphene sheets.\cite{dassarma:2007}
Magnetic properties in bulk graphene in particular occur in a regime where large screening effects
(allowed by a dispersive kinetic energy) minimize the impact of the long-range intra-band Coulomb interaction between electrons
(See, e.g., Refs.~\onlinecite{esquinazi:2003,peres:2005,herbut:2006,ohldag:2007,sheehy:2007,huang:2009,cervenka:2009}).  Flat
kinetic energy bands, by contrast, do not allow screening and therefore strongly emphasize interaction effects by default.
Furthermore, conventional perturbative
treatments of the interaction (in comparison to the kinetic energy)
fail in flat-bands due to the absence of a small parameter.

Most theoretical studies of interactions in flat bands use the
Hubbard model with an on-site term.\cite{nagaoka:1966,tasaki:1998a,mielke:1999}  The on-site
Hubbard model incorporates just the energy penalty for two electrons
to occupy the same site while ignoring the long range part of the
Coulomb interaction.  The on-site term leads to surprising ground
states in the flat-band Hubbard model.  For example, work by Nagoaka
\cite{nagaoka:1966} finds ferromagnetism in flat bands at specific
fillings, near one particle per site.  This is in stark contrast to
antiferromagnetism favored by super exchange in dispersive bands.

Graphene edges, nanoribbons, and dots present physical systems
hosting flat bands.  Theoretical modeling typically relies on the
on-site Hubbard model to make predictions.  For example, work
studying flat bands in on-site Hubbard models of zig-zag nanoribbons
\cite{fujita:1996,yazyev:2011} uses meanfield theory to argue for
ferromagnetic states along nanoribbon edges but antiferromagnetic
coupling between edges. An ab initio calculation \cite{lee:2005} and
a work using both the weak-coupling renormalization group and the
density-matrix renormalization-group calculation
\cite{hikibara:2003} provide similar results.

\begin{figure}[t]
\centerline{\includegraphics [width=3 in] {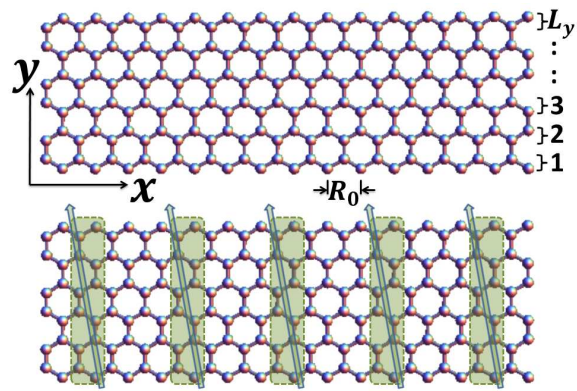}} \caption{(Color
online) Top: Schematic of a zig-zag nanoribbon of carbon atoms.
Bottom:  Schematic of a ferromagnetic crystal with one electron for
every three unit cells in one band.  The shaded areas correspond to
single unit cells and the arrows indicate aligned electron spins.}
\label{ribbon}
\end{figure}

Motivated by recent experiments on graphene
nanoribbons,\cite{tao:2011} we construct interacting lattice models
of electrons in flat-band nanoribbons.  We focus on zig-zag
nanoribbons because here, in contrast to arm-chair ribbons, two flat
bands arise near the Fermi level even in the absence of adsorbates.
\cite{nakada:1996}  In the top panel of Fig.~\ref{ribbon}, we
schematically show a zig-zag nanoribbon where $R_0$ ($\sim 2.46
\mathrm{{\AA}}$) labels the width of a unit cell along the ribbon
($x$ direction) and $L_y$ labels the number of zig-zag chains across
the ribbon ($y$ direction). At low densities the absence of
intra-band screening in flat bands suggests that the long-range part
of the Coulomb interaction is relevant.  We therefore construct
models that include even the long-range part of the interaction.  We choose to
model a very specific regime: flat-bands in zig-zag nanoribbons,
because we expect the absence of conventional screening to cause flat-band electrons to order in a way which is
completely distinct from electrons in bulk graphene.

The goal of our work is to establish a set of working Hamiltonians
of zig-zag nanoribbons.  We construct a single-particle basis of
Wannier functions.  We use our basis to compute the interaction
matrix elements.  We then establish a projection protocol that sets
up approximate flat-band models.  Projection into flat bands
delocalizes basis states due to quantum interference. The resulting
flat-band models are highly non-trivial (incorporating two bands,
long-range interactions, and spin) and can lead to many quantum
ground states even in the absence of significant dispersion.  We
make simple estimates of the low energy properties of our models at
odd denominator fillings of a single band.

We argue that, at low densities, the long-range part of the Coulomb
interaction supports ferromagnetic quantum crystals (bottom panel of
Fig.~\ref{ribbon}).  Crystalline order projected into the flat band
incorporates quantum superpositions because basis states delocalize.
At low fillings direct spin exchange leads to an effective
Heisenberg model.  Our simple estimates therefore predict
ferromagnetic crystalline order in certain parameter regimes.  Our
work sets the stage for more accurate studies of our models with a
general class of Jastrow-correlated wavefunctions that apply to flat
bands. \cite{wang:2011}

Our protocol differs from conventional band-structure calculations.
Flat bands, in contrast to dispersive bands, are, by default,
strongly interacting.  Conventional applications of density
functional theory accurately model the effect of core electrons
while making very local approximations for the Coulomb interaction
between mobile electrons.  Flat bands require accurate treatment of
the long-range portion of the unscreened Coulomb interaction between
otherwise mobile electrons.

In Section~\ref{singleparticle} we consider the band structure that
arises from non-interacting tight-binding models of zig-zag
nanoribbons. Two flat bands are identified. In
Section~\ref{wanniersection} we construct localized single-particle
basis states, orthonormal Wannier functions, from carbon $\pi_{z}$
orbitals in the honeycomb lattice model of zig-zag nanoribbons.
Sections~\ref{oneband} and ~\ref{twoband} use the Wannier functions
to explicitly compute Coulomb interaction matrix elements for one
and two flat bands, respectively. Section~\ref{projection} defines a
projection scheme which limits the total many-body model to the
flat-band portion of the single-particle spectrum.
Section~\ref{lowenergy} sorts terms in the many-body model to argue
that, at low fillings, energetics favor ferromagnetic quantum
crystals. Section~\ref{summary} summarizes and looks forward to more accurate studies of the models constructed here.\\

\section{Flat Bands in Zig-Zag Graphene Nanoribbons}
\label{singleparticle}

We consider interacting electrons hopping among carbon sites forming
zig-zag graphene nanoribbons (Fig.~\ref{ribbon}). We first model the
electrons in a simple non-interacting tight-binding picture.  The
single-particle tight-binding Hamiltonian is: \cite{castroneto:2009}
\begin{eqnarray}
H_{0}=-t\sum_{\langle n,m\rangle}(\hat{c}_{n}^{\dagger}\hat{c}_{m}^{\vphantom{\dagger}}+\text{h.c.}),
\label{H0}
\end{eqnarray}
where the hopping integral is $t\sim 2.7$ eV for graphene
\cite{castroneto:2009} and the sum is along bonds of the honeycomb
lattice.  The second-quantized operator $\hat{c}_{n}^{\dagger}$
creates a fermion at a site $n$.  Labels $n$ and $m$ indicate
lattice sites, in contrast to labels for unit cells, $i,j,k,l$, used
in the following.

\begin{figure}[t]
\centerline{\includegraphics [width=3 in] {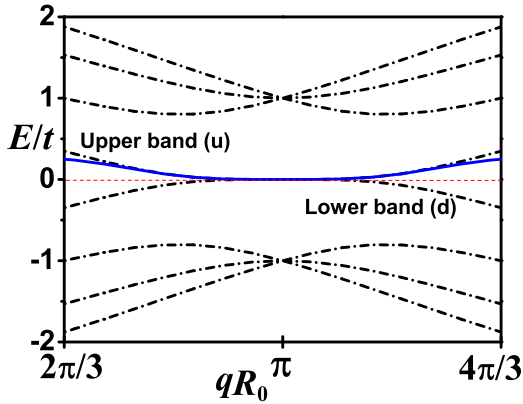}} \caption{(Color
online) The dot-dashed lines indicate the energy eigenvalues of
Eq.~(\ref{H0}) versus wavevector for a nanoribbon of width $L_{y}=4$
and a $q$-space mesh of $N=44$.  The solid line shows the
approximate expression for the energy, Eq.~(\ref{dispersion}).  Two
flat bands form near $qR_{0}=\pi$.  In the large $L_{y}$ limit, the
bands flatten for $2\pi/3\leq qR_{0}\leq4\pi/3$. } \label{band}
\end{figure}

Two bands near the Fermi level flatten for large ribbon widths.
\cite{nakada:1996}  An example band structure for a narrow width,
$L_y=4$, is shown in Fig.~\ref{band}. Near the fermi surface, the
conduction band (upper band, $u$) and valence band (lower band, $d$)
are nearly degenerate for wavevectors $q$ in the region $qR_0 \in
[2\pi/3,4\pi/3]$ and form flat bands.  For larger widths the bands
flatten considerably.

We examine the band width with simple ansatz flat-band
single-particle states. \cite{nakada:1996}  Considering states in
the region $qR_0 \in [2\pi/3,4\pi/3]$ with even $L_y$:
\begin{eqnarray}
\phi_{\pm}(q,y)&=&(\phi_{A}(q,y),\pm\phi_{B}(q,y))^{T} \nonumber \\
&=&((-u_{q})^{y-1},\pm(-1)^{y-1}(u_{q})^{L_y-y})^{T},
\label{disper1}
\end{eqnarray}
for $y=1,...,L_y$ where $u_{q}\equiv2\mathrm{cos}(qR_0/2)$, the
energy dispersion in band $\Gamma=u,d$ can be computed analytically:
\begin{eqnarray}
|E_{\Gamma}(q)|&\approx&|\phi(q,y)^{T}H_{0}(q)\phi(q,y)|/|\phi(q,y)|^2 \nonumber\\
&=&t(1-u_{q}^2)u_{q}^{L_y}/(1-u_{q}^{2L_y}),
\label{dispersion}
\end{eqnarray}
with
\begin{equation}
H_{0}(q)=t\left(\begin{array}{cc}
0 & Q(q) \\
Q^{\dagger}(q) & 0
\end{array}\right),
Q(q)=\left(\begin{array}{cccc}
u_{q} & 0 & .. & 0 \\
1 & u_{q} & 0 & .. \\
: & : & : & : \\
0 & .. & 1 & u_{q}
\end{array}\right).
\nonumber
\end{equation}
Figure~\ref{band} compares Eq.~(\ref{dispersion}) with the exact results from Eq.~(\ref{H0}).

Eq.~(\ref{dispersion}) can be used to determine the bandwidth.  For
partially filled lattices a narrow range of single-particle basis
states will be occupied. The bandwidth for states in the flat-band
sector vanishes for ribbons with large width:
\begin{eqnarray}
\vert E_{\Gamma}(|q-\pi|\rightarrow\pi/3)\vert\rightarrow \frac{t}{L_{y}}.
\end{eqnarray}
From this estimate we see that band dispersion plays a small role
for dilute ribbons with increasing ribbon widths.

A vanishing bandwidth, due to quantum interference, leaves the
interaction as the dominant term in the many-body Hamiltonian for
electrons.   For dilute ribbons we will work in the approximation
that $H_{0}$ adds an overall constant energy shift to the spectrum.
The full Hamiltonian adds the unscreened Coulomb interaction:
\begin{eqnarray}
H_{\text{total}}=H_{0}+H_{V}.
\label{htotal}
\end{eqnarray}
In the following we treat the dispersion as a small correction  to
the interacting term.  We project the Hamiltonian into the basis of
flat-band states.   Our model becomes:
\begin{eqnarray}
H_{\text{total}}&=&\sum_{\mathbf{q}\in \mathrm{BZ},\sigma,\Gamma}E_{\Gamma}(q)\hat{c}_{\mathbf{q}\sigma\Gamma}^{\dag}\hat{c}_{\mathbf{q}\sigma\Gamma}^{\vphantom{\dagger}}
+H_V\nonumber\\
&\rightarrow&\text{constant}+\mathcal{P}^{\dagger}_{\mathrm{FB}}H_V\mathcal{P}_{\mathrm{FB}}^{\vphantom{\dagger}},
\label{eqnprojection}
\end{eqnarray}
where the first equality is written in terms of the creation (annihilation) operators $\hat{c}_{\mathbf{q}\sigma\Gamma}^{\dag}$ ($\hat{c}_{\mathbf{q}\sigma\Gamma}^{\vphantom{\dagger}}$) for Bloch states at wavevector $q$ and band $\Gamma$ in the Brillouin zone (BZ), which are related to the operators for single-particle basis states by a Fourier transform:
\begin{eqnarray}
\hat{c}_{j\sigma\Gamma}^{\dag}=\frac{1}{\sqrt{N}}\sum_{\mathbf{q}\in \mathrm{BZ}}e^{i\mathbf{q}\cdot\mathbf{R}_j}\hat{c}_{\mathbf{q}\sigma\Gamma}^{\dag}.
\label{creator}
\end{eqnarray}
Here $\mathbf{R}_j$ is the lattice vector of the $j$th unit cell,
$N$ defines the number of unit cells and $q$-space mesh, and
$\sigma\in\{\uparrow, \downarrow\}$ denotes spin.
$\mathcal{P}^{\dagger}_{\mathrm{FB}}$ denotes projection into flat
bands such that the many-body eigenstates are constructed from Bloch
states with $qR_0 \in [2\pi/3, 4\pi/3]$.  Many-body states
incorporating these values of $q$ will have essentially no kinetic
energy. We consider this model as a centerpiece to understanding the
electronic properties of flat-band nanoribbons at low densities.

To explore possible many-body states in zig-zag nanoribbons we
construct an accurate form for Eq.~(\ref{eqnprojection}) in the
flat-band basis.  We note that the absence of any dispersion
excludes intra-band screening as in ordinary Fermi liquids.  Thus
many-body eigenstates are determined entirely by the interplay
between various terms in the interaction.  It is therefore crucial
to accurately determine the interacting terms in
Eq.~(\ref{eqnprojection}) as prescribed by our choice of
single-particle basis.  To construct an accurate single-particle
basis we revisit the underlying simple tight-binding model formed
from overlapping $\pi_{z}$ orbitals.  We construct orthonormal
Wannier functions from these orbitals.  The Wannier functions will
serve as single-particle basis states, allowing the construction of
competing terms in a many-body model.

\section{Single-Particle Basis States: Flat-Band Wannier Functions}
\label{wanniersection}

In this section we construct a set of single-particle basis states
in nanoribbon flat bands.  We superpose carbon $\pi_z$ orbitals to
form orthogonal Wannier functions.  The Wannier functions will then,
in later sections, be used to accurately determine interaction
matrix elements.

\begin{figure}[t]
\centerline{\includegraphics [width=3 in] {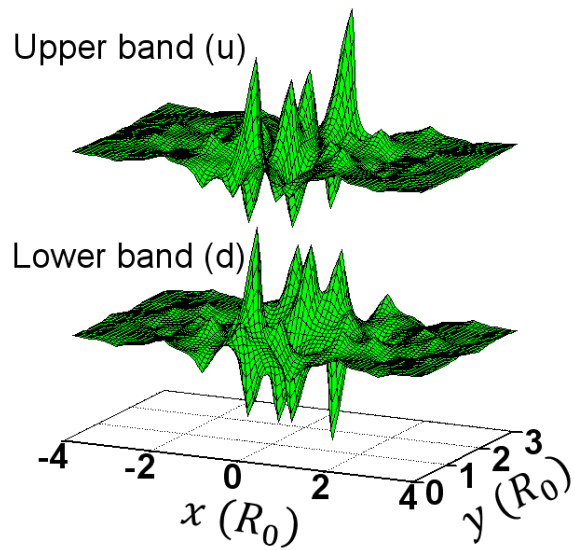}} \caption{(Color
online) Two-dimensional Wannier functions plotted as a function of
position in the lattice for a ribbon of width $L_{y}=4$.  The
Wannier functions tend to localize near the ribbon edges.}
 \label{wannier}
\end{figure}

In an isolated band the Wannier functions are given by:
\begin{eqnarray}
 W_j(\mathbf{r})=W_0(\mathbf{r}-\mathbf{R}_j)=\frac{V}{(2\pi)^D}\oint_{\mathrm{BZ}}d\mathbf{q}e^{-i\mathbf{q}\cdot\mathbf{R}_j}\Psi_{\mathbf{q}}(\mathbf{r}),
\label{eq1}
\end{eqnarray}
where $D$ is the dimension,  $V$ is the volume of unit cell. The
Bloch functions are
$\Psi_{\mathbf{q}}(\mathbf{r})=\sum_{m=1}^{M}C_{mq}\chi_{mq}(\mathbf{r})$,
with $M$ atomic sites per unit cell.

To make contact with first principles calculations on graphene
nanoribbons \cite{castroneto:2009} we form Bloch functions from
carbon $\pi_z$ orbitals, $\phi(\mathbf{r})=\sqrt{\xi^5/\pi}ze^{-\xi
r}$. The basis states become
$\chi_{mq}(\mathbf{r})=(1/\sqrt{N})\sum_{j=0}^{N-1}e^{i\mathbf{q}\cdot\mathbf{R}_{j}}\phi(\mathbf{r}-\mathbf{R}_{j}-\mathbf{T}_m)$,
where $\mathbf{T}_m$ is the location of the $m$th atom in the unit
cell.

The coefficients $C_{mq}$ and energy eigenvalues $E(q)$ are obtained
from diagonalization of the secular equation:
\begin{eqnarray}
\left[ \tilde{O}^{-1} \tilde{H}(q) \right] {\bf C}_{q}=E(q){\bf C}_{q},
\label{matrixequation}
\end{eqnarray}
where the matrix $\tilde{H}$ follows from the tight-binding
Hamiltonian $H_0$: $\tilde{H}(q)_{mn}=\int
d\mathbf{r}\chi_{mq}^{*}(\mathbf{r}) H_0 \chi_{nq}(\mathbf{r})$  and
the elements of the overlap matrix $\tilde{O}$ are given by
$O_{mn}=\int
d\mathbf{r}\chi_{mq}^{*}(\mathbf{r})\chi_{nq}(\mathbf{r})$.  The
eigenvectors ${\bf C}_{q}\equiv\{C_{1q},...,C_{Mq}\}^{T}$ yield the
coefficients used in the definition of the Wannier functions.  In
the tight-binding approximation we set $O_{mn}$ proportional to the
elements of the identity matrix, $\delta_{mn}$.

\begin{figure}[t]
\centerline{\includegraphics [width=3 in] {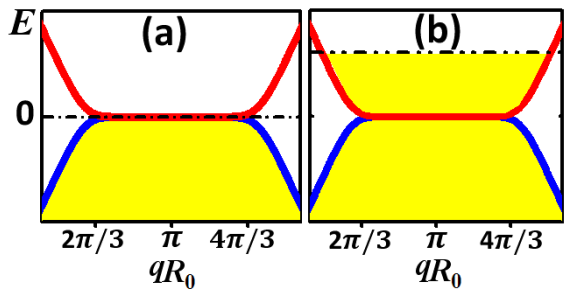}} \caption{(Color
online) Left panel: Schematic of the energy dispersion for a wide
ribbon with the Fermi level between the degenerate energy bands $u$
and $d$.  In this regime low lattice filling allows us to accurately
ignore the finite dispersion near the band edges.  Right panel:  The
same as the left panel but at larger fillings of the upper $u$ band.
Here the flat-band approximation will only be a good approximation
if the Coulomb interaction is much larger than the band width.  }
\label{Ek}
\end{figure}

We solve Eq.~(\ref{matrixequation}) to construct orthonormal Wannier
functions. We consider a one-dimensional lattice of unit cells along
the nanoribbon.  The discrete wavevectors become $\mathbf{q}=(2\pi
q/NR_0) \hat{\mathbf{x}}$.  The Wannier function located at
$\mathbf{R}_j$ is then:
\begin{eqnarray}
 W_j(\mathbf{r})=\frac{1}{N}\sum_{q=0}^{N-1}e^{-i2\pi qj/N}\Psi_{q}(\mathbf{r}).
\label{eq3}
\end{eqnarray}
The Wannier functions defined in this way are unique for a $D=1$
single band model \cite{kohn:1959} but for higher dimensions and
with more bands they are not necessarily unique. \cite{marzari:1997}
We choose a specific set of single-particle basis states by
enforcing $C_{mq}=|C_{mq}|$ at the edge atomic site $m=1$.  As a
result we obtain a set of real Wannier functions symmetric about the
$x$ axis.

The above Wannier function can be written as a summation over all
local atomic orbitals $\phi(\mathbf{r})$ located at sites
$\mathbf{r}_{mi}=\mathbf{T}_m+\mathbf{R}_{i}$.  Rewriting $W$ at the
origin gives:
\begin{eqnarray}
 W_0(\mathbf{r})=N_f \sum_{m=1}^{M}\sum_{i=0}^{N-1}\alpha_{mi}\phi(\mathbf{r}-\mathbf{r}_{mi}),
\label{eq4}
\end{eqnarray}
with weights $\alpha_{mj}=\sum_{q=0}^{N-1}C_{mq}e^{i2\pi qj/N}$ and
normalization constant $N_f$.  The coefficients $\alpha$ completely
determine our choice of basis.

We can extend our calculation of the Wannier functions to include
both the upper and lower bands. A denser sampling in momentum space
(i.e., larger $N$) yields more accurate Wannier functions. In
practice, we find that the Wannier function has already converged
when taking $N=44$ for $L_{y}=4$. The Wannier functions of upper and
lower bands for the same sample ribbon are shown in
Fig.~\ref{wannier}.  We note that the Wannier functions localize
symmetrically about $x=0$ with an extension of less than four unit
cells.  The Wannier functions are also symmetric (antisymmetric)
along $y$ for the upper (lower) band.

The flat-band Wannier functions constructed here correspond to a
specific choice of single-particle basis.  By constructing
superpositions of these functions we can equivalently construct a
model using basis states localized on either edge of the ribbon via
a simple rotation in the two-band space.  Viewed in this way our
model implicitly includes inter-edge coupling in narrow ribbons
because we work in the basis of $u$ and $d$ bands as opposed to a
two-edge basis.

Our approach can be used to model graphene edges.  Our study applies to the edge states
of very wide ribbons provided we superpose our $u$ and $d$ band Wannier functions to construct
left and right edge Wannier functions.  Our model can then be used to study edges of very wide
ribbons.  But we stress that our model \emph{cannot} apply to the electrons
in the center of graphene because we have considered bands in nanoribbons that carry over only to edge states
in the wide ribbon limit (For a discussion see Ref.~\onlinecite{nakada:1996}).  In what follows we focus on
narrow ribbons and only consider Wannier functions in the $u$ and $d$ band
basis.

\section{One-Band Coulomb Model}
\label{oneband}

Interaction effects determine the low energy properties of
Eq.~(\ref{htotal}) in the absence of significant dispersion.  When
the chemical potential lies between the nearly flat bands of zig-zag
nanoribbons, the Coulomb interaction sets the dominant energy scale
and mitigates response.  Figure~\ref{Ek} shows schematic band
structures for a wide ribbon with the chemical potential at the band
degeneracy (left) and far from the flat-band region (right).  In
what follows we focus on dilute systems corresponding to the left
panel.  We can, as a first approximation, assume that the valence
band is inert and that only the conduction band, $u$, will be active
under external probes.  Projection into the flat $u$ band implies
that the Coulomb interaction alone operates in the massively
degenerate subspace formed from $u$ band single-particle basis
states.  In this section we will consider the $u$ band only.  In the
following section we will construct a model of both the $u$ and $d$
bands.

We consider an unscreened Coulomb interaction in a single band:
\begin{eqnarray}
\sum_{i,j,k,l,\sigma\sigma^{'}}\mathcal{V}_{ijkl}\hat{c}_{i\sigma}^\dag \hat{c}_{j\sigma^{'}}^\dag \hat{c}_{k\sigma^{'}}^{\vphantom{\dagger}}\hat{c}_{l\sigma}^{\vphantom{\dagger}},
\end{eqnarray}
where the second-quantized operators $\hat{c}_{i\sigma}^{\dag}$
($\hat{c}_{i\sigma}^{\vphantom{\dagger}}$) create (annihilate) a
fermion with spin $\sigma$ in a Wannier state centered at the $i$th
unit cell.  The matrix elements $\mathcal{V}$ depend on the basis.
We can rewrite the Coulomb interaction in the $u$ band in a
suggestive form:
\begin{eqnarray}
H_V^{u} &=&V_0\sum_{i} n_{i\uparrow}n_{i\downarrow}+\sum_{i<j}V_{ij}n_in_j-\sum_{i< j}J_{ij}\mathbf{S}_i\cdot\mathbf{S}_j\nonumber\\
&+&\frac{1}{2}\sum_{\{i,j\}\nsubseteq\{k,l\},\sigma\sigma^{'}}V_{ijkl}\hat{c}_{i\sigma}^\dag \hat{c}_{j\sigma^{'}}^\dag \hat{c}_{k\sigma^{'}}^{\vphantom{\dagger}}\hat{c}_{l\sigma}^{\vphantom{\dagger}}.
\label{onebandH}
\end{eqnarray}
Here, the single-component and total density operators are
$n_{i\sigma}=\hat{c}_{i\sigma}^{\dag}\hat{c}_{i\sigma}^{\vphantom{\dagger}}$
and $n_i=n_{i\uparrow}+n_{i\downarrow}$, respectively. The spin
operators
$\mathbf{S}_i=(1/2)\sum_{\sigma\sigma^{'}}\hat{c}_{i\sigma}^{\dag}${\boldmath$\tilde{\sigma}$}$_{\sigma\sigma^{'}}\hat{c}_{i\sigma^{'}}^{\vphantom{\dagger}}$
are defined in terms of the Pauli matrices
{\boldmath$\tilde{\sigma}$}.

Eq.~(\ref{onebandH}) keeps all terms in the full Coulomb
interaction.  We compute the matrix elements in the basis of Wannier
functions in the $u$ band.  Integral equations for the coefficients
are given in the appendix, Eqs.~(\ref{twobandcoefficients}).  The
first term is the ordinary single-site Hubbard term which is the
only term that is commonly used in models of flat-band nanoribbons
(See, e.g., Refs.~\onlinecite{yazyev:2011} and
\onlinecite{fujita:1996}).  The second term captures the diagonal
portion of the Coulomb interaction at long range.  The absence of a
dispersion implies that these terms can be relevant and must be kept
in accurate models, especially at low fillings.  The third term, the
direct exchange term, favors ferromagnetism for $J_{ij} >0$.   The
last term represents remaining off-diagonal terms due to the Coulomb
interaction.  We find, by direct calculation, that the last terms
are very small compared to the other terms for a single band.

We compute coefficients in Eq.~(\ref{onebandH}) explicitly.  We
perform the integrals in Eqs.~(\ref{twobandcoefficients}) by
approximating the exponential part of the $\pi_z$ orbital,
$\phi(\mathbf{r})$, as a linear combination of three Gaussian
functions:
$\sum_{s}\gamma_s(128\beta_s^5/\pi^3)^{1/4}ze^{-\beta_sr^2}$.  We
obtain the parameters $\gamma_s$ and $\beta_s$ from the STO-3G
package. \cite{emsl} Data for fitting the $\pi_z$ orbital with
$\xi=1.72$ are listed in Table \ref{tab1}. For numerical results
shown here and in the following sections, we use the Bohr radius,
$a_0=0.53 \mathrm{{\AA}}$, as the unit of length and the Coulomb
energy $e^2/4\pi\epsilon a_0$ ($\sim$  27.2 eV in vacuum) as the
unit of energy.
\begin{table}[t]
\caption{Fitting parameters for the Gaussian approximation to the
$\pi_z$ orbital with $\xi=1.72$.} \centering
\begin{tabular}{|l|c|c|c|}
  \hline
  $s$            & 1          & 2          & 3 \\[-.1ex] \hline
  $\gamma_s$     & 0.15591627 & 0.60768372 & 0.39195739 \\[-.1ex]
  $\beta_s$      & 2.9412494  & 0.6834831  & 0.2222899  \\
  \hline
\end{tabular}
\label{tab1}
\end{table}

\begin{table}[t]
\caption{Matrix elements for one-band (u band) case for inter-unit
cell separations of up to $4R_{0}$. } \centering
\begin{tabular}{|c|c|c|c|c|}
  \hline
  \multicolumn{5}{|l|}{$V_0$=2.24$\times10^{-1}$} \\
  \hline
  $|i-j|$  & 1                   & 2                   & 3                   & 4                   \\
  $J_{ij}$ & 2.34$\times10^{-2}$ & 4.68$\times10^{-3}$ & 9.21$\times10^{-4}$ & 1.69$\times10^{-4}$ \\
  $V_{ij}$ & 1.43$\times10^{-1}$ & 9.56$\times10^{-2}$ & 6.83$\times10^{-2}$ & 5.23$\times10^{-2}$ \\
  \hline
\end{tabular}
\label{tab2}
\end{table}

Table~\ref{tab2} lists the coefficients computed for an $L_y=4$
ribbon.  As we see, all coefficients are positive and can be sorted
by $V_0 > V_{ij} > J_{ij} > 0$.  The ground state can be determined
by an interplay between leading terms in Eq.~(\ref{onebandH}) and
the chemical potential. These coefficients suggest that partially
filled single bands support the formation of ferromagnetic crystals.
However, the large Coulomb interaction may cause mixing between the
$u$ and $d$ bands.  In the next section we construct a two-band
model.

\section{Two-Band Coulomb Model}
\label{twoband}

We now consider a more comprehensive two-band model.  The $u$ and
$d$ bands in the flat-band region are essentially degenerate for
wide ribbon widths.  The Coulomb interaction can in principle favor
occupancy of both bands or the occupancy of a single band.  Accurate
estimates of coefficients in the full two-band model will allow
exploration of the two-band energy landscape to determine the band
occupancy in future work.

We construct Wannier functions in both the $u$ and $d$ bands.  The
Hamiltonian is dominated by the following terms:
\begin{eqnarray}
H_V^{ud}&=&\sum_{i,\Gamma}V_{0}^{\Gamma}n_{i\Gamma\uparrow}n_{i\Gamma\downarrow}\nonumber\\
&+&\sum_{i}\left (V_{ii}^{'}n_{iu}n_{id}-J_{ii}^{'}\mathbf{S}_{iu}\cdot\mathbf{S}_{id}\right )\nonumber \\
 &+&\sum_{i<j,\Gamma}(V_{ij}^{\Gamma}n_{i\Gamma}n_{j\Gamma}-J_{ij}^{\Gamma}\mathbf{S}_{i\Gamma}\cdot\mathbf{S}_{j\Gamma})\nonumber\\
 &+&\sum_{i<j}\sum_{\Gamma \neq \Gamma^{'}}(V_{ij}^{'}n_{i\Gamma}n_{j\Gamma^{'}}-J_{ij}^{'}\mathbf{S}_{i\Gamma}\cdot\mathbf{S}_{j\Gamma^{'}})\nonumber\\
 &+&\sum_{i<j}\sum_{\Gamma \neq \Gamma^{'}}\sum_{\sigma\sigma^{'}}(
  V_{ij}^{''}\hat{c}_{i\Gamma\sigma}^{\dag}\hat{c}_{j\Gamma^{'}\sigma^{'}}^{\dag}\hat{c}_{j\Gamma\sigma^{'}}^{\vphantom{\dagger}}\hat{c}_{i\Gamma^{'}\sigma}^{\vphantom{\dagger}}\nonumber\\
&+&V_{ij}^{'''}\hat{c}_{i\Gamma\sigma}^{\dag}\hat{c}_{j\Gamma^{'}\sigma^{'}}^{\dag}\hat{c}_{i\Gamma^{'}\sigma^{'}}^{\vphantom{\dagger}}\hat{c}_{j\Gamma\sigma}^{\vphantom{\dagger}}).
\label{fullmodel}
\end{eqnarray}
We have checked, by direct calculation, that other terms involving
three and four centers are much smaller than terms kept in Eq.
(\ref{fullmodel}). Here we see the Hubbard and ferromagnetic terms
as in the one-band case. The last term indicates a non-trivial band
exchange term. The integrals for all coefficients are listed in the
Appendix.

\begin{table}[t!]
\caption{Matrix elements for the two-band case with $L_y=4$ for
inter-unit cell separations of up to $4R_{0}$. } \centering
\begin{tabular}{|c|c|c|c|c|c|}
  \hline
  \multicolumn{3}{|l}{$V^{d}_{0}$=2.28$\times10^{-1}$}  & \multicolumn{3}{l|}{$V^{u}_{0}$=2.24$\times10^{-1}$}\\
  \multicolumn{3}{|l}{$V^{'}_{ii}$=1.91$\times10^{-1}$} & \multicolumn{3}{l|}{$J_{ii}^{'}$=1.32$\times10^{-1}$}\\
  \hline
  $|i-j|$        & 1                   & 2                   & 3                   & 4                   &  $D_w$ \\
  $V_{ij}^{d}$   & 1.44$\times10^{-1}$ & 9.51$\times10^{-2}$ & 6.79$\times10^{-2}$ & 5.21$\times10^{-2}$ & 1.01 \\
  $V_{ij}^{u}$   & 1.43$\times10^{-1}$ & 9.56$\times10^{-2}$ & 6.83$\times10^{-2}$ & 5.23$\times10^{-2}$ & 1.02 \\
  $V_{ij}^{'}$   & 1.46$\times10^{-1}$ & 9.56$\times10^{-2}$ & 6.81$\times10^{-2}$ & 5.22$\times10^{-2}$ & 1.02 \\
  $J_{ij}^{d}$   & 2.60$\times10^{-2}$ & 3.04$\times10^{-3}$ & 5.75$\times10^{-4}$ & 1.09$\times10^{-4}$ & \\
  $J_{ij}^{u}$   & 2.34$\times10^{-2}$ & 4.68$\times10^{-3}$ & 9.21$\times10^{-4}$ & 1.69$\times10^{-4}$ & \\
  $J_{ij}^{'}$   & 1.62$\times10^{-2}$ & 3.14$\times10^{-3}$ & 6.54$\times10^{-4}$ & 1.27$\times10^{-4}$ & \\
  $V_{ij}^{''}$  & 2.06$\times10^{-2}$ & 7.44$\times10^{-3}$ & 2.95$\times10^{-3}$ & 1.35$\times10^{-3}$ & \\
  $V_{ij}^{'''}$ & 1.05$\times10^{-2}$ & 1.82$\times10^{-3}$ & 3.49$\times10^{-4}$ & 6.43$\times10^{-5}$ & \\
  \hline
\end{tabular}
\label{tablely4}
\end{table}

Eq.~(\ref{fullmodel}) presents a central result of our work.  The
two-band model must be studied for different fillings and different
widths to determine expected ground states.  Tables~\ref{tablely4}
and ~\ref{tablely10} show numerically computed coefficients for two
example widths, $L_{y}=4$ and $10$.

The tables show that the electron configurations are determined
primarily by the diagonal components of the Coulomb interaction
(rows 1-3).  These rows are nearly equal indicating a band symmetry,
as expected.  These rows govern the charge degrees of freedom.  Rows
4-6 govern the spin degrees of freedom.  The positive elements
support ferromagnetism.  The last two rows give rise to band
exchange effects.

\begin{figure}[t!]
\centerline{\includegraphics [width=3 in] {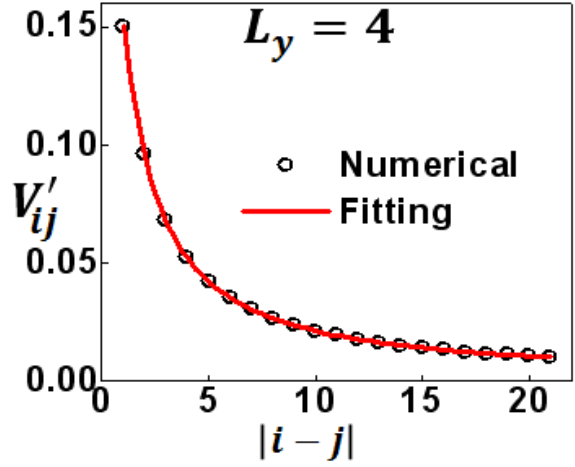}} \caption{The
diagonal component of the inter-band Coulomb interaction
($V_{ij}^{'}$) for a zigzag graphene nanoribbon with $L_{y}=4$ and
$N=44$. Circles are from numerical evaluation of
Eqs.~(\ref{twobandcoefficients}).  The solid line is a fit with
Eq.~(\ref{fitting}) and $D_w=1.02$.} \label{fittingw4}
\end{figure}

\begin{figure}[t !]
\centerline{\includegraphics [width=3 in] {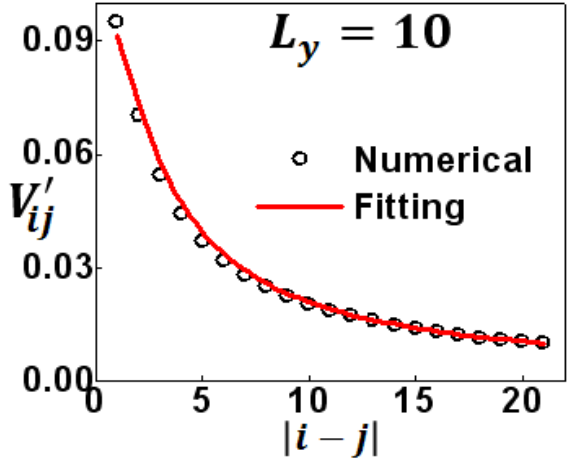}} \caption{The same
as Fig.~\ref{fittingw4} but for $L_{y}=10$ with $D_w=2.12$.}
\label{fittingw10}
\end{figure}

We construct a  simple fitting form for the first three rows.  We
note that the coefficients $V_{ij}^{\Gamma}$ and $V_{ij}^{'}$ can be
thought of as a softened Coulomb interaction between smeared charges
located at separate unit cells $i$ and $j$.  For large separations
the charges appear as point charges and interact through the Coulomb
interaction but at short ranges our basis states smear the electron
charge over the width of the ribbon.  We approximate
$V_{ij}^{\Gamma}$ and $V_{ij}^{'}$ with a convenient analytic form:
\begin{equation}
V_{i\neq j}\approx \left (\frac{e^2}{4 \pi \epsilon a_0}\right)\frac{a_{0}/R_{0}}{\sqrt{|i-j|^2+D_w^2}},
\label{fitting}
\end{equation}
where the  fitting parameter $D_w$ is dependent on the width of the
ribbon and can be determined with a numerical fitting as shown in
Figs.~\ref{fittingw4} and \ref{fittingw10}. The last column of
Tables~\ref{tablely4} and ~\ref{tablely10} shows $D_{w}$ obtained by
fitting.

Eq.~(\ref{fitting}) can  be used to approximate the coefficients in
Eq.~(\ref{fullmodel}) at low filling.  At low filling
$V_{ij}^{\Gamma}$ and $V_{ij}^{'}$ determine the configuration of
charges.  It then suffices to consider spin exchange terms at the
separations fixed by $V_{ij}^{\Gamma}$ and $V_{ij}^{'}$.  We use
this procedure to suggest possible low energy solutions to
Eq.~(\ref{fullmodel}).

\section{Flat-Band Projection}
\label{projection}

The flat-band  limit, Eq.~(\ref{eqnprojection}), establishes a
unique set of non-perturbative models.  In this section we construct
a set of operators that allow flat-band projection of models
constructed in the previous sections.  In the following section we
will then use the projected models in simple estimates of the low
energy physics.

\begin{table}[t!]
\caption{The same as Table~\ref{tablely4} but for $L_y=10$.}
\centering
\begin{tabular}{|c|c|c|c|c|c|}
  \hline
  \multicolumn{3}{|l}{$V^{d}_{0}$=1.21$\times10^{-1}$}  & \multicolumn{3}{l|}{$V^{u}_{0}$=1.17$\times10^{-1}$}\\
  \multicolumn{3}{|l}{$V_{ii}^{'}$=1.03$\times10^{-1}$} & \multicolumn{3}{l|}{$J_{ii}^{'}$=5.90$\times10^{-2}$}\\
  \hline
  $|i-j|$        & 1                   & 2                   & 3                   & 4                   & $D_w$ \\
  $V_{ij}^{d}$   & 8.93$\times10^{-2}$ & 6.96$\times10^{-2}$ & 5.47$\times10^{-2}$ & 4.45$\times10^{-2}$ & 2.09  \\
  $V_{ij}^{u}$   & 8.74$\times10^{-2}$ & 6.84$\times10^{-2}$ & 5.41$\times10^{-2}$ & 4.41$\times10^{-2}$ & 2.15  \\
  $V_{ij}^{'}$   & 9.08$\times10^{-2}$ & 6.93$\times10^{-2}$ & 5.44$\times10^{-2}$ & 4.43$\times10^{-2}$ & 2.12  \\
  $J_{ij}^{d}$   & 2.82$\times10^{-2}$ & 4.80$\times10^{-3}$ & 1.10$\times10^{-3}$ & 3.36$\times10^{-4}$ & \\
  $J_{ij}^{u}$   & 2.65$\times10^{-2}$ & 6.99$\times10^{-3}$ & 1.51$\times10^{-3}$ & 4.30$\times10^{-4}$ & \\
  $J_{ij}^{'}$   & 1.65$\times10^{-2}$ & 4.68$\times10^{-3}$ & 1.15$\times10^{-3}$ & 3.10$\times10^{-4}$ & \\
  $V_{ij}^{''}$  & 1.75$\times10^{-2}$ & 1.00$\times10^{-2}$ & 5.82$\times10^{-3}$ & 3.52$\times10^{-3}$ & \\
  $V_{ij}^{'''}$ & 1.28$\times10^{-2}$ & 2.85$\times10^{-3}$ & 6.30$\times10^{-4}$ & 1.85$\times10^{-4}$ & \\
  \hline
\end{tabular}
\label{tablely10}
\end{table}

To enforce  flat-band projection we limit all $q$-space sums to the
flat-band region (FBR) $qR_0 \in [2\pi/3,4\pi/3]$.  We can therefore
project into a single band by considering a flat-band operator that
limits itself to the FBR:
\begin{eqnarray}
\hat{b}_{j\sigma}^{\dag}\equiv\frac{1}{N}\sum_{l}\sum_{\mathbf{q}\in \mathrm{FBR}}e^{i\mathbf{q}\cdot(\mathbf{R}_j-\mathbf{R}_l)}\hat{c}_{l\sigma}^{\dag}.
\label{proj-creator}
\end{eqnarray}
This operator  creates states centered around the unit cell at
$\mathbf{R}_j$.  We note that the states created by this operator
have finite overlap with neighbors at $\mathbf{R}_{j+1}$ when the
flat-band region does not encompass the entire Brillouin zone.  In
the limit that the flat band encompasses the entire Brillouin zone
the overlap between neighboring states vanishes and we have
$\hat{b}_{j\sigma}^{\dag}\rightarrow\hat{c}_{j\sigma}^{\dag}$. Thus,
the projection into a flat band that incorporates only a fraction of
the Brillouin zone delocalizes basis states.

We can  rewrite our model in terms of projected density and spin
operators.  The single-component and total projected density
operators are $\rho_{i\sigma}\equiv
\hat{b}_{i\sigma}^{\dag}\hat{b}_{i\sigma}^{\vphantom{\dagger}}$ and
$\rho_i\equiv\rho_{i\uparrow}+\rho_{i\downarrow}$, respectively. The
projected spin operators are defined as:
\begin{eqnarray}
\slashed {\mathbf S }_{j} \equiv \frac{1}{2N}\sum_{\sigma\sigma^{'}}\sum_{\mathbf{q},\mathbf{q'}\in \mathrm{FBR}}e^{i(\mathbf{q}-\mathbf{q'})\cdot\mathbf{R}_j}\hat{c}_{\mathbf{q}\sigma}^{\dag}{\boldsymbol {\tilde{\sigma}}}_{\sigma\sigma^{'}}\hat{c}_{\mathbf{q'}\sigma^{'}}^{\vphantom{\dagger}}.
\label{proj-spin}
\end{eqnarray}
We  stress that the projected operators do \emph{not} exhibit
ordinary commutation relations because the underlying operators
create overlapping states, i.e., $\langle 0 \vert \hat{b}_{j+1}
\hat{b}_{j}^{\dag}|0\rangle\neq0$.

The  projected Hamiltonian can be rewritten entirely in terms of the
above projected operators. Starting from an unprojected model, we
impose projection using the following replacements: $c\rightarrow
b,n\rightarrow\rho,$ and ${\mathbf S}\rightarrow\slashed {\mathbf
S}$.  For example, the flat-band projected Coulomb interaction in
the $u$ band becomes:
\begin{eqnarray}
\mathcal{P}^{\dagger}_{u}H^{u}_V\mathcal{P}_{u}^{\vphantom{\dagger}}&=&V_0\sum_{i} \rho_{i\uparrow}\rho_{i\downarrow}+\sum_{i<j}V_{ij}\rho_i\rho_j-\sum_{i< j}J_{ij}\slashed {\mathbf S}_i\cdot\slashed {\mathbf S}_j\nonumber\\
&+&\frac{1}{2}\sum_{\{i,j\}\nsubseteq\{k,l\},\sigma\sigma^{'}}V_{ijkl}\hat{b}_{i\sigma}^\dag \hat{b}_{j\sigma^{'}}^\dag \hat{b}_{k\sigma^{'}}^{\vphantom{\dagger}}\hat{b}_{l\sigma}^{\vphantom{\dagger}}.
\label{proj-onebandH}
\end{eqnarray}
The  projected two-band model can also be obtained with a similar
replacement applied to $H^{ud}_V$.

\section{Low Energy Properties}
\label{lowenergy}

We use  flat-band projection to discuss possible low energy states
of Eq.~(\ref{eqnprojection}) based on simple energetic arguments.  A
detailed quantitative analysis of low energy states is beyond the
scope of the present work.  We make progress by ordering terms
according to dominant energy scales.  We then focus on example
lattice fillings.

To consider  low energy solutions of Eq.~(\ref{eqnprojection}) we
first examine the kinetic term.  The kinetic term enforces a
flat-band projection provided the chemical potentials lies near the
flat band, i.e., Fig.~\ref{Ek}a.  It is then sufficient to require
that many-body eigenstates of $H_{V}$ utilize Bloch states with
$qR_0 \in [2\pi/3, 4\pi/3]$.  We can analyze Eq.~(\ref{fullmodel})
with this $q$-space restriction by using projected operators
constructed in the previous section.

We first point  out an intrinsic energetic ordering to each of the
terms in Eq.~(\ref{fullmodel}).  We rewrite each of the terms
according to an approximate ordering by energy and in the projected
space:
\begin{eqnarray}
\mathcal{P}^{\dagger}_{ud}H_V^{ud}\mathcal{P}^{\vphantom{\dagger}}_{ud}&=&\sum_{i,\Gamma}V_{0}^{\Gamma}\rho_{i\Gamma\uparrow}\rho_{i\Gamma\downarrow} \nonumber \\
 &+&\sum_{i, j,\Gamma,\Gamma'}\left(\overline{V}_{ij}^{\Gamma,\Gamma'}\rho_{i\Gamma}\rho_{j\Gamma'}-\overline{J}_{ij}^{\Gamma,\Gamma'}\slashed {\mathbf S}_{i\Gamma}\cdot\slashed {\mathbf S}_{j\Gamma'}\right)\nonumber\\
 &+&H_{\text{Band-exch}},
\label{Vsimplified}
\end{eqnarray}
where we  have redefined the diagonal Coulomb terms:
$\overline{V}_{i<j}^{\Gamma\neq\Gamma'}\equiv V'_{ij}$,
$\overline{V}_{ii}^{\Gamma=d,\Gamma'=u}\equiv V'_{ii}$, and
$\overline{V}_{i<j}^{\Gamma=\Gamma'}\equiv V^{\Gamma}_{ij}$,
otherwise $\overline{V}_{ij}^{\Gamma,\Gamma'}=0$. (Note that our
direct calculations find
$\overline{V}_{ij}^{\Gamma\neq\Gamma'}\approx
\overline{V}_{ij}^{\Gamma=\Gamma'}$.) We have also redefined the
off-diagonal exchange terms:
$\overline{J}_{i<j}^{\Gamma\neq\Gamma'}\equiv J'_{ij}$,
$\overline{J}_{ii}^{\Gamma=d,\Gamma'=u}\equiv J'_{ii}$, and
$\overline{J}_{i<j}^{\Gamma=\Gamma'}\equiv J^{\Gamma}_{ij}$,
otherwise $\overline{J}_{ij}^{\Gamma,\Gamma'}=0$.  The last term in
Eq.~(\ref{Vsimplified}) corresponds to the last term in
Eq.~(\ref{fullmodel}).

We can  understand the low energy properties of the first three
terms in Eq.~(\ref{Vsimplified}) at a few specific fillings.
Considering an inert $d$ band, we assume that the $u$ band is
partially filled at odd denominators, $\nu_{u}=1/(2p+1)$, where
$p=1,2,...$. ($\nu$ indicates the number of particles per basis
state.)  Ignoring $H_{\text{Band-exch}}$ allows a decomposition of
basis states into the $u$ and $d$ bands.  An inert $d$ band implies
that the inter-band interaction leads to an overall shift of the
chemical potential.  A strong external gate bias canceling this
shift should be able to maintain the $u$-band filling
$\nu_{u}=1/(2p+1)$.

In the  limit of commuting projected density operators it is well
known \cite{hubbard:1978} that the first terms in
Eq.~(\ref{Vsimplified}) lead to a charge order, i.e.,
one-dimensional Wigner crystals with lattice spacing $2p+1$. We
therefore expect that the $u$-band electrons form a classical Wigner
crystal in the limit that the flat band encompasses the entire
Brillouin zone.  The bottom panel of Fig.~\ref{ribbon} depicts a
classical crystal configuration in a single spin state.

In the limit  that the projected density operators do not commute,
the case for zig-zag nanoribbons, we predict quantum crystals in
partially filled bands.  Quantum crystals arise, in direct analogy
to Wigner crystals, as eigenstates of the projected density
operators.  For example, a trial quantum crystal state at
$\nu_{u}=1/(2p+1)$ in spin state $\sigma$ is given by:
\begin{eqnarray}
\prod_{j=0}\hat{b}_{2pj+j,\sigma u}^{\dag}\vert 0\rangle.
\end{eqnarray}
This trial  state appears to minimize the energy of the first two
terms in Eq.~(\ref{Vsimplified}) by separating flat-band charges by
an average of $2p$ unit cells.  Thus the first two terms in
Eq.~(\ref{Vsimplified}) impose a rigid charge order in the $u$ band.
However, the charges are significantly delocalized.  A finite
overlap among neighbors implies that the charges exist in a
superposition of several different unit cells at once: a quantum
crystal.

Provided a  rigid charge ordering we consider the next lowest energy
scale: low energy spin properties of Eq.~(\ref{Vsimplified}).  We
approximate the spin-spin coupling with an effective Heisenberg
model for the u-band particles at $\nu_{u}=1/(2p+1)$:
\begin{eqnarray}
H_{\text{eff}}^{p}=-J^{u}_{0,2p+1}\sum_{i}\slashed{\mathbf S}_{i,u}\cdot \slashed{\mathbf S}_{i+2p+1,u}.
\label{Heff}
\end{eqnarray}
Eq.~(\ref{Heff}) applies  to the case of a single band at odd
denominator filling.

The ground states  of Eq.~(\ref{Heff}) are ferromagnetic quantum
crystals. The low energy spin excitations are ferromagnetic magnons.
The underlying rigid charge order enforces a large magnon
wavelength.  At $\nu_{u}=1/(2p+1)$ spin wave theory yields
excitation energies:
\begin{eqnarray}
\hbar \omega_{q}=2J^{u}_{0,2p+1}  \left[1-\cos((2p+1)R_{0}q) \right].
\end{eqnarray}
This dispersion  offers a clear indicator of ferromagnetic crystals
in the spin degrees of freedom.

At finite temperatures  the Mermin-Wagner theorem asserts that
spin-spin correlations decay with a finite length scale in the
one-dimensional Heisenberg model. \cite{mermin:1966}  Thus,
ferromagnetic ordering holds only up to small length scales.  The
spin-spin correlation length at non-zero temperatures, $T$, for the
Heisenberg chain with exchange coupling $J$ is: \cite{kopietz:1989}
\begin{eqnarray}
\frac{\xi_{T}}{(2p+1)R_{0}}=\frac{AJ}{4T}\left[1+B(8T/J)^{1/2}/\pi+\mathcal{O}\left(\frac{T}{J}\right)\right],
\end{eqnarray}
where $A\approx1.1$ and $B\approx0.65$.  Our  results suggest that
for $L_{y}=10$ at $T=1K$ with $J^{u}_{i,i+3}\approx 1.5\times10^{-3}
(e^{2}/4\pi\epsilon_{0}a_{0})\approx 473K$ the correlation length is
$\xi_{T} /(2p+1)R_{0}\approx 133$.  Thus about 390 unit cells
containing 130 $u$-band electrons are included in the formation of a
fully magnetized domain at $\nu_{u}=1/3$ for these parameters.

\section{Summary and Outlook}
\label{summary}

We constructed interacting flat-band lattice models of zig-zag
nanoribbons.  A single-particle basis of orthonormal Wannier
functions were built from carbon $\pi_{z}$ orbitals in a
honeycomb-ribbon lattice.  The single-particle basis was used to
explicitly compute the Coulomb matrix elements for two ribbon
widths, $L_{y}=4$ and $10$.  The total model, Eqs.~(\ref{htotal})
and (\ref{fullmodel}), was then projected into the flat bands of the
single-particle spectrum.  The projected flat-band model,
Eq.~(\ref{Vsimplified}), suggests ferromagnetic quantum crystal
ground states.

Our flat-band  model, Eq.~(\ref{Vsimplified}), sets the stage for
more accurate analyses with a combination of numerics and many-body
wavefunctions.  The absence of a small parameter calls for a
combination of variational studies and diagonalization to verify
proposed ground and excited states.\cite{wang:2011}  In addition to
crystals discussed here, uniform quantum liquids are also
possible.\cite{wang:2011}

The models  constructed here focus on key physics of interacting
flat bands but exclude several realistic effects.  In experiments on
graphene nanostructures many corrections may be required before
making a detailed comparison with experiment.  For example, edge
roughness, defects, and substrate disorder can destroy the flat-band
approximation.  Furthermore, inter-band screening has also been
ignored in the current study.  While intra-band screening was
implicitly incorporated in our model,  screening from nearby bands
could lead to corrections to the pure Coulomb model studied here,
e.g., RKKY-type interactions. \cite{brey:2007,saremi:2007}

\vspace{0.2cm}
\section{Acknowledgements}

We thank the Thomas F. Jeffress and Kate Miller Jeffress Memorial Trust, Grant No. J-992, for support.\\

\section{Appendix}

The coefficients in Eqs.~\ref{onebandH} and \ref{fullmodel} are given by:
\begin{eqnarray}
V_{0}^{\Gamma}&=&\int\frac{d^2\mathbf{r}d^2\mathbf{r'}}{|\mathbf{r}-\mathbf{r'}|}|W_{0\Gamma}(\mathbf{r})W_{0\Gamma}(\mathbf{r'})|^2, \nonumber\\
J_{ij}^{\Gamma}&=&2\int\frac{d^2\mathbf{r}d^2\mathbf{r'}}{|\mathbf{r}-\mathbf{r'}|}W_{i\Gamma}^{*}(\mathbf{r})W_{j\Gamma}(\mathbf{r})W_{i\Gamma}(\mathbf{r'})W_{j\Gamma}^{*}(\mathbf{r'}),  \nonumber\\
V_{ij}^{\Gamma}&=&\int\frac{d^2\mathbf{r}d^2\mathbf{r'}}{|\mathbf{r}-\mathbf{r'}|}|W_{i\Gamma}(\mathbf{r})W_{j\Gamma}(\mathbf{r'})|^2-\frac{1}{4}J_{ij}^{\Gamma},  \nonumber\\
J_{ij}^{'}&=&2\int\frac{d^2\mathbf{r}d^2\mathbf{r'}}{|\mathbf{r}-\mathbf{r'}|}W_{iu}^{*}(\mathbf{r})W_{jd}(\mathbf{r})W_{iu}(\mathbf{r'})W_{jd}^{*}(\mathbf{r'}),  \nonumber\\
V_{ij}^{'}&=&\int\frac{d^2\mathbf{r}d^2\mathbf{r'}}{|\mathbf{r}-\mathbf{r'}|}|W_{iu}(\mathbf{r})W_{jd}(\mathbf{r'})|^2-\frac{1}{4}J_{ij}^{'},  \nonumber\\
V_{ij}^{''}&=&\int\frac{d^2\mathbf{r}d^2\mathbf{r'}}{|\mathbf{r}-\mathbf{r'}|}W_{iu}^{*}(\mathbf{r})W_{id}(\mathbf{r})W_{ju}(\mathbf{r'})W_{jd}^{*}(\mathbf{r'}),  \nonumber\\
V_{ij}^{'''}&=&\int\frac{d^2\mathbf{r}d^2\mathbf{r'}}{|\mathbf{r}-\mathbf{r'}|}W_{iu}^{*}(\mathbf{r})W_{ju}(\mathbf{r})W_{id}(\mathbf{r'})W_{jd}^{*}(\mathbf{r'}),  \nonumber\\
V_{ijkl}&=&\int\frac{d^2\mathbf{r}d^2{\mathbf{r}}^{'}}{|\mathbf{r}-{\mathbf{r}}^{'}|}W_{iu}^{*}(\mathbf{r})W_{lu}(\mathbf{r})W_{ju}^{*}({\mathbf{r}}^{'})W_{ku}({\mathbf{r}}^{'}).
\label{twobandcoefficients}
\end{eqnarray}
The last term is used only in Eq.~(\ref{onebandH}).


\begin{thebibliography}{99}

\bibitem{novoselov:2004} K. S. Novoselov, A. K. Geim, S. V. Morozov, D. Jiang, Y. Zhang, S. V. Dubonos, I. V. Grigorieva, and A. A. Firsov, Science {\bf 306}, 666 (2004).

\bibitem{castroneto:2009} A. H. Castro Neto, F. Guinea, N. M. R. Peres, K. S. Novoselov, and A. K. Geim,
Rev. Mod. Phys. {\bf 81}, 109 (2009).

\bibitem{nakada:1996} K. Nakada, M. Fujita, G. Dresselhaus, and M. S. Dresselhaus, Phys. Rev. B {\bf 54}, 17954 (1996).

\bibitem{lin:2009} H. H. Lin, T. Hikihara, H.T. Jeng, B. L. Huang, C. Y. Mou, and X. Hu, Phys. Rev. B {\bf 79}, 035405 (2009).

\bibitem{potasz:2010} P. Potasz, A. D. G\"{u}\c{c}l\"{u}, and P. Hawrylak, Phys. Rev. B {\bf 82}, 075425 (2010).

\bibitem{kusakabe:2003}  K. Kusakabe and M. Maruyama,
Phys. Rev. B {\bf 67}, 092406 (2003).

\bibitem{ezawa:2007}
M. Ezawa, Phys. Rev. B {\bf 76}, 245415 (2007); J.
Fern\'{a}ndez-Rossier and J. J. Palacios, Phys. Rev. Lett. {\bf 99},
177204 (2007); A. D. G\"{u}\c{c}l\"{u}, P. Potasz, O. Voznyy, M.
Korkusinski, and P. Hawrylak, \emph{ibid.} {\bf 103}, 246805 (2009).

\bibitem{vanevic:2009}
M. Vanevi\'{c}, V. M. Stojanovi\'{c}, and M. Kindermann, Phys. Rev.
B {\bf 80}, 045410 (2009); J. A. F\"{u}rst, T. G. Pedersen, M.
Brandbyge, and A. P. Jauho, \emph{ibid.}  {\bf 80}, 115117 (2009).

\bibitem{dassarma:2007} S. Das Sarma, E. H. Hwang, and Wang-Kong Tse, Phys. Rev. B {\bf 75}, 121406R (2007); S. Das Sarma, S. Adam, E. H. Hwang, and E. Rossi, Rev. Mod. Phys. {\bf 83}, 407 (2011).

\bibitem{esquinazi:2003} P. Esquinazi, D. Spemann, R. H\"{o}hne, A. Setzer, K. H. Han, and T. Butz,
Phys. Rev. Lett. {\bf 91}, 227201 (2003).

\bibitem{peres:2005} N. M. R. Peres, F. Guinea, and A. H. Castro Neto,
Phys. Rev. B {\bf 72}, 174406 (2005).

\bibitem{herbut:2006} I. F. Herbut,
Phys. Rev. Lett. {\bf 97}, 146401 (2006).

\bibitem{ohldag:2007} H. Ohldag, T. Tyliszczak, R. H\"{o}hne, D. Spemann, P. Esquinazi, M. Ungureanu, and T. Butz,
Phys. Rev. Lett. {\bf 98}, 187204 (2007).

\bibitem{sheehy:2007} D. E. Sheehy and J. Schmalian,
Phys. Rev. Lett. {\bf 99}, 226803 (2007).

\bibitem{huang:2009} B. L. Huang and C. Y. Mou,
Euro. Phys. Lett. {\bf 88}, 68005 (2009);
B.L. Huang, M.C. Chang, and C. Y. Mou,
Phys. Rev. B {\bf82}, 155462 (2010).

\bibitem{cervenka:2009}  J. \v{C}ervenka, M. I. Katsnelson and C. F. J. Flipse,
Nature Phys. {\bf 5}, 840 (2009).

\bibitem{nagaoka:1966} Y. Nagoaka,
Phys. Rev. {\bf 147}, 392 (1966).

\bibitem{tasaki:1998a} H. Tasaki,
Prog. Theor. Phys. {\bf 99}, 489 (1998).

\bibitem{mielke:1999} A. Mielke,
Phys. Rev. Lett. {\bf 82}, 4312 (1999).

\bibitem{yazyev:2011} O. V. Yazyev,  R. B. Capaz, and S. G. Louie,
Phys. Rev. B {\bf 84}, 115406 (2011).

\bibitem{fujita:1996} M. Fujita, K. Wakabayashi, K. Nakada, and K. Kusakabe,
J. Phys. Soc. Jpn. {\bf 65}, 1920 (1996).

\bibitem{lee:2005} H. Lee, Y. W. Son, N. Park, S. Han, and J. Yu,
Phys. Rev. B {\bf 72}, 174431 (2005).

\bibitem{hikibara:2003} T. Hikihara, X. Hu, H. H. Lin, and C. Y.
Mou, Phys. Rev. B {\bf 68}, 035432 (2003).

\bibitem{tao:2011} C. Tao, L. Jiao, O. V. Yazyev, Y. C. Chen, J. Feng, X. Zhang, R. B. Capaz, J. M. Tour, A. Zettl, S. G. Louie, H. Dai, and M. F. Crommie, Nature Phys. {\bf 7}, 616 (2011).

\bibitem{wang:2011} Hao Wang and V.W. Scarola,
Phys. Rev. B {\bf 83}, 245109 (2011).

\bibitem{kohn:1959} W. Kohn,
Phys. Rev. {\bf 115}, 809 (1959).

\bibitem{marzari:1997} N. Marzari and D. Vanderbilt,
Phys. Rev. B {\bf 56}, 12847 (1997).

\bibitem{emsl} EMSL  Basis Set Exchange Library v1.2.2 at http://bse.pnl.gov/bse/.

\bibitem{hubbard:1978} J. Hubbard,
Phys. Rev. B {\bf 17}, 494 (1978).

\bibitem{mermin:1966} N. D. Mermin and H. Wagner,
Phys. Rev. Lett. {\bf 17}, 1133 (1966).

\bibitem{kopietz:1989} P. Kopietz,
Phys. Rev. B {\bf 40}, 5194 (1989).

\bibitem{brey:2007} L. Brey, H. A. Fertig, and S. Das Sarma, Phys. Rev.
Lett. {\bf 99}, 116802 (2007).

\bibitem{saremi:2007} S. Saremi,
Phys. Rev. B {\bf 76}, 184430 (2007).

\end{thebibliography}
\end{document}